\documentclass[
  aps,
  pre,
  twocolumn,
  superscriptaddress,
  groupedaddress
]{revtex4}

\usepackage{amsmath}    
\usepackage{graphicx}   
\usepackage{verbatim}   
\usepackage[usenames]{color}  
\usepackage{subfigure}  
\usepackage{hyperref}   
\usepackage{amssymb}
\usepackage{epstopdf}
\usepackage{float}

\begin{document}

\title{RNA topology remolds electrostatic stabilization of viruses}
 \author{Gonca Erdemci-Tandogan}
 \author{Jef Wagner}
 \affiliation{Department of Physics and Astronomy,
   University of California, Riverside, California 92521, USA}
 \author{Paul van der Schoot}
 \affiliation{Group Theory of Polymers and Soft Matter, Eindhoven University of Technology, P.O. Box 513, 5600 MB Eindhoven,
   The Netherlands}
 \affiliation{ Institute for Theoretical Physics,
   Utrecht University,
   Leuvenlaan 4, 3584 CE Utrecht, The Netherlands}
 \author{Rudolf Podgornik}
 \affiliation{Department of Theoretical Physics, J. Stefan Institute, SI-1000 Ljubljana, Slovenia}
 \affiliation{Department of Physics, University of Ljubljana, SI-1000 Ljubljana, Slovenia}
 \author{Roya Zandi}
 \affiliation{Department of Physics and Astronomy,
   University of California, Riverside, California 92521, USA}

\begin{abstract}
Simple RNA viruses efficiently encapsulate their genome into a
nano-sized protein shell: the capsid. Spontaneous co-assembly of
the genome and the capsid proteins is driven predominantly by electrostatic interactions
between the negatively charged RNA and the positively charged
inner capsid wall.  Using field theoretic formulation we show that
the inherently branched RNA secondary structure allows viruses to {\sl maximize}
the amount of encapsulated genome and make assembly more efficient, allowing viral RNAs to out-compete
cellular RNAs during replication in infected host cells.
\end{abstract}

\maketitle

\section{INTRODUCTION}
\vspace{-0.25cm}

Simple viruses encapsulate their genetic material into a
protein shell, measuring no more than about 15 nm
across for the smallest viruses and about 28 nm for a typical
(plant) virus \cite{Bancroft,bruinsma}. Under many circumstances, {\sl in vitro} virus assembly is spontaneous and driven primarily
by electrostatic interactions between negative charges on the backbone
of the polynucleotide, usually single-stranded (ss) RNA, and positive
charges on the virus coat proteins \cite{Cornelissen2007,Ren2006,Bogdan,Vanderschoot2007,Anze2,DanielDragnea2010,Zlotnick,Hsiang-Ku}. 
However, recent experiments indicate that RNA plays a role that goes beyond its
polyelectrolyte (PE) nature, as some RNAs are encapsulated more efficiently
than others \cite{Comas}. For example, when viral RNA1 of BMV
(brome mosaic virus) and CCMV (cowpea chlorotic mottle virus) are
mixed in solution with the capsid proteins from CCMV, the BMV RNA is
packaged three times more efficiently \cite{Comas}. As the two RNAs differ
in the amount of branching and their tertiary structure, both a
straightforward consequence of their primary sequence,
there must be a tight connection between capsid packing preferences
and the structure of RNA \cite{Yoffe2008,Li-tai}.

In many viruses the number of negative charges on the ssRNA is larger than the
number of positive charges on the virus coat proteins
\cite{Shklovskii,Vanderschoot,Belyi,Ting}. This overcharging phenomenon is intriguing and has been the subject of many papers.
Belyi and Muthukumar examined a sample of actual viruses and found the ratio of the
RNA charge to structural charge on the inner capsid surface to be around
$1.6$ \cite{Belyi}. While it seems to be feasible to
encapsulate linear polymers with a number of charges as much as nine
times that on the capsid proteins \cite{Chuck2008},
which would result in strong ``overcharging'' of the virion,
recent experiments show that the {\it optimal} number
of charges residing on the linear PE is \textit{less} than the total
number of charges on the inner surface of viral shells \cite{Cadena2011}, implying
``undercharging'' of the complex of capsid proteins with {\it linear} polyelectrolytes.  This naturally leads to the question of which
RNA features are implicated in the overcharging of the virion.

In what follows we show that RNA secondary structural features,  such as branching,
have a pronounced effect on the genome encapsulation capacity and thus could explain the phenomena of overcharging observed 
in viral particles.

In virtually all theoretical studies published to date investigating the
overcharging in viruses, the genome is modeled as a simple linear
polyelectrolyte chain \cite{Vanderschoot,Siber,Hagan2010,Hagan2013}. 
{Thus the phenomenon of overcharging is associated with many factors other than the structure of RNA
\cite{Belyi,Shklovskii,Vanderschoot,Ting,Siber}.
Our numerical solutions of the polyelectrolyte Poisson-Boltzmann theory, without any additional assumptions 
regarding the effective monomer charge, do not support a universal overcharging of the virion.  
In fact, we find that mean-field PE theories for linear polymer chains lead to an undercharging phenomenon with 
fewer negative charges on the chain than positive ones on the capsid. This is consistent with several other 
recent numerical studies \cite{Siber,Ting,Bogdan}. The overcharging in the viral particles has been explained 
through the genome-capsid N-terminal interactions
and/or Manning condensation \cite{Belyi,Shklovskii,Vanderschoot,Siber}. However, it is important to note that the phenomenon of 
overcharging is also observed  in viruses in which the number of charges on N-terminals is not significant, {\it e.g.}, ~Dengue 
and yellow fever viruses \cite{Shklovskii}.

While the theoretical studies of linear polymers shed some light on the overcharging phenomenon, 
recent experiments reveal the importance of RNA structure that goes beyond its polyelectrolyte nature as a
linear charged chain \cite{Comas,Yoffe2008,Stockley2012,Stockley2013}. 
\rm Intra-chain base paring, promoted by hydrogen bonding between mutually complementary
nucleotides along the backbone, leads to a highly branched structure of the RNA
molecule that furthermore promotes its compaction in free solution.

\begin{figure}
  \includegraphics[width=8.0cm,height=5cm]{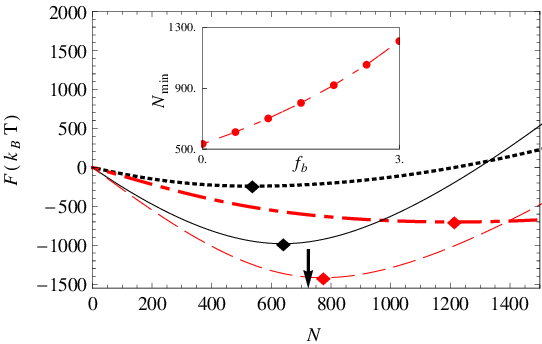}
  \caption{\label{fenergy}Encapsulation free energy as a function of
    monomer numbers for a linear polymer with $f_b$=0 (solid and dotted lines)
      and a branched polymer with $f_b$=3 (dashed and dot-dashed lines) at
    two different values of $\mu$, corresponding to salt concentrations 10 $mM$ (solid and dashed
    lines) and 100 $mM$ (dotted and dotted-dashed
    lines). The arrow
      indicates the monomer number at which the full virus particle
      (capsid + polyelectrolyte) becomes neutral. Inset shows
    the position of the minimum $N_{\text{min}}$ vs. the branching
    fugacity $f_b$ for 100 $mM$ salt concentration. Other
    parameters used are $\upsilon=0.5$, $\tau=-1$, $\sigma=0.4$, $b=12$,
    and $T=300 K$, typical for RNA and virus capsids \cite{bruinsma,Vanderschoot2009,Anze,Vanderschoot}.}
\end{figure}

In this paper, we present a model that takes into account the
combined effect of RNA branching and the genome-capsid protein electrostatic
interactions. We find that the larger the inherent propensity to form
branch points quantified by the fugacity, $f_b$, the larger is the
optimal chain length that can be accommodated in the capsid,
consistent with {\it in vitro} experiments. These results are evident
from Fig.~\ref{fenergy}, showing the displacement in the position of the
encapsulation free energy minimum towards longer chains as the
branching fugacity increases. The inset to Fig.~\ref{fenergy}, showing the position of the
minimum $N_{\text{min}}$ as a function of the propensity for branching
$f_b$, directly demonstrates this effect.

The figure also illustrates that the encapsulation free energy becomes more negative with increasing
propensity of RNA to form branch points for a given number of monomers.
This stabilization behavior suggests that branching is not only
conducive to more efficient packing of the genome material into the virus shell, but also
allows viral RNAs that have more branch points than other types of cellular RNAs \cite{Yoffe2008} to
out-compete the latter during replication in infected, susceptible host cells.

\section{THEORY}
\vspace{-0.25cm}

To obtain the optimal length and the free energy associated with the
encapsulated RNA inside a capsid, we model RNA as a generic flexible
branched polyelectrolyte.  Because of the physical character of the
base pairing, the degree of branching of RNAs is statistical and may
in the process of encapsulation be affected by interaction with the
charges located on or near the inner surface of the protein coat.  To
this end, we consider only the case of annealed branched polymers in
this letter \cite{McPherson}.

Further, we consider that the RNA interacts with positive
charges residing on the inner surface of a sphere, where for simplicity
we additionally presume that the charges are not localized but smeared
out uniformly. For a large proportion of viruses the positive charges are
indeed located on the inner surface of the capsid that in reality is
not a perfect sphere but has a structure on the nanometer scale  \cite{Anze}.
For some viruses positively charged disordered domains on the coat
proteins point into the capsid cavity in a brush-like fashion \cite{Shklovskii},
a feature that we do not include in our coarse-grained model at this stage.

In the mean-field, ground-state approximation, the free energy of a
negatively charged polymer chain confined to a positively charged,
infinitely thin spherical shell can be written as
\begin{multline} \label{free_energy}
  F = \!\!\int\!\! {\mathrm{d}^3}{{{r}}}\Big[
    \tfrac{1}{6} |{\nabla\Psi({\bf{r}})}|^2
    +W\big[\Psi({\bf{r}}) \big]\\
    -\tfrac{1}{8 \pi \lambda_B} |{\nabla\Phi({\bf{r}})}|^2
    -2\mu\cosh\big[ \Phi({\bf{r}})\big]
    + \tau \Phi({\bf{r}})\Psi^2({\bf{r}})
    \Big]\\
  + \int\!\! {\mathrm{d}^2}r \Big[ \sigma \, \Phi({\bf{r}}) \Big].
\end{multline}
All quantities in Eq.~\ref{free_energy} are dimensionless, so energies are in units of thermal
energy $k_BT$ and lengths in units of the statistical step length
(Kuhn length) of the polymer, $a$. Here, $\tau$ denotes the linear
charge density of the polymer, $\sigma$ the surface charge density of
the shell, $\Psi^2({\bf{r}})$ the monomer density at
position $\mathbf{r}$, and $\Phi({\bf{r}})$ the mean electrostatic
potential. 
The parameter $\mu$ is the fugacity (density) of the monovalent salt ions, and
corresponds to the concentration of salt ions in the bulk. The (dimensionless) Bjerrum length, $\lambda_B$,
is a measure of the dielectric constant of the solvent, corresponding to about
$0.7$ nm for water at room temperature. The square gradient term in the
first line of Eq.~\eqref{free_energy} describes the entropic cost for
a non-uniform polymer density, and the last two lines in Eq.~\eqref{free_energy} describe the
electrostatic interactions between the polyelectrolyte, the salt ions,
and the charged capsid at the level of Poisson-Boltzmann theory
\cite{Siber}. {{The full derivation of the standard form of the free energy can be found in Refs.
\cite{Borukhov,Borukhov1,Borukhov2,Borukhov3,Shafir}. 
In addition to the standard form, we add}} the $W[\Psi]$ term that describes the
statistics of an annealed branched polymer \cite{Lubensky,Nguyen-Bruinsma,Lee-Nguyen,Elleuch}, given
explicitly by
\begin{align} \label{W_branched}
  W[\Psi]&=\frac{1}{2}\upsilon \Psi^4 -f_e\Psi-\frac{f_b}{6} \Psi^3,
\end{align}
where $\upsilon$ is the (dimensionless) excluded volume and $f_e$ and
$f_b$ are the fugacities of the end- and branch-points, respectively.

In our description, the stem-loop or hair-pin configurations in RNA
structures are counted as end points. The number of end and branch
points $N_e$ and $N_b$ of the polymer depend on the fugacities $f_e$ and $f_b$ through
\begin{align}\label{NeNb}
N_e =- f_e \frac{\partial{F}}{\partial{f_e}} \qquad {\rm and} \qquad N_b =- f_b \frac{\partial{F}}{\partial{f_b}}.
\end{align}
Since we consider only the case
of a single encapsulated polymer with no closed loops,
there is a constraint on the number of end and branch points,
\begin{equation}\label{branch_constraint}
  N_e = N_b+2,
\end{equation}
with the degree of branching controlled by the
fugacity of branch points $f_b$.  The chain is linear if $f_b=0$, and
becomes more branched as $f_b$ increases. The fugacity of
endpoints $f_e$ is not a free parameter in the system, it is set through the above constraint,
Eq.~\eqref{branch_constraint}. In addition, the total number of
polyelectrolyte monomers inside the capsid is fixed \cite{deGennes,Hone}, \textit{i.e.},
\begin{equation}\label{constraint}
  N= \int {\mathrm{d}^3}{r} \; \Psi^2 ({\bf{r}}),	
\end{equation}
which we enforce by introducing a Lagrange multiplier, $E$, when
minimizing the free energy.

We obtain the polyelectrolyte profile and electrostatic potential by
varying the free energy functional with respect to fields $\Psi(r)$
and $\Phi(r)$ {{\cite{SiberZandi2010}}}. The resulting coupled set of non-linear equations describes
the monomer density field, $\Psi$, and the electrostatic potential
$\Phi_{in}$, in the interior of the capsid, and the usual
Poisson-Boltzmann equation for the electrostatic potential,
$\Phi_{out}$, in the exterior of the capsid, {\sl viz.},
\begin{subequations} \label{euler}
  \begin{align}
    \frac{1}{6} \nabla^2 \Psi (\mathbf{r})&
    =-E {\Psi} (\mathbf{r}) + \tau \Phi_{in}(\mathbf{r}) \Psi(\mathbf{r})+
    \frac{1}{2} \frac{\partial W}{\partial \Psi} \label{euler_a}\\
    \nabla^2 \Phi_{in} (\mathbf{r}) &
    = \tfrac{1}{\lambda_D ^{2}}
    \sinh \big [ \Phi_{in}(\mathbf{r}) \big ]
    -  \tfrac{\tau}{{2 \lambda_D ^{2}}  \mu}
    {\Psi}^2 (\mathbf{r})\label{euler_b}\\
    \nabla^2 \Phi_{out} (\mathbf{r}) &
    = \tfrac{1}{\lambda_D ^{2}}
    \sinh \big [ \Phi_{out}(\mathbf{r}) \big ] \label{euler_c}
  \end{align}
\end{subequations}
where $\lambda_D=1/\sqrt{8\pi \lambda_B \mu}$ is the (dimensionless)
Debye screening length. The polymer segment concentration outside the
capsid is assumed to be zero, $\Psi = 0$. Equations \eqref{euler} along
with the constraints in Eqs.~\eqref{branch_constraint} and \eqref{constraint} represent a set of coupled nonlinear
differential equations that, subject to appropriate boundary
conditions, can only be solved numerically for the unknown parameters $f_e$ and $E$ and fields $\Psi$ and $\Phi$.

Assuming that the positive surface charge density, $\sigma$, is fixed, the
electrostatic boundary condition (BC) is obtained by minimizing the free
energy with respect to $\Phi$ on the surface, ${\hat n.}{\nabla \Phi_{in}}-{\hat n.}{\nabla \Phi_{out}} =
{{4\pi\lambda_B}}{\sigma}$.  The choice of boundary
conditions for $\Psi$ depends on how the polymer interacts with the
capsid surface through non-electrostatic forces. The strong short-ranged
repulsion (as would be the case if we had included an excluded volume
term between the polyelectrolyte monomers and the capsid proteins)
leads to Dirichlet BCs. However, it turns out that for the
large surface charge densities relevant to viruses, our conclusions are robust
and do not depend on the choice of BC; {we come back to this below}. 
In this paper, we focus on Neumann BCs, directly obtained from the minimization of the free energy 
in Eq.~\ref{free_energy} with respect to the polymer density field, $\psi$, on the surface.

\section{RESULTS}
\vspace{-0.25cm}

\begin{figure}
  \includegraphics[width=8.0cm,height=5cm]{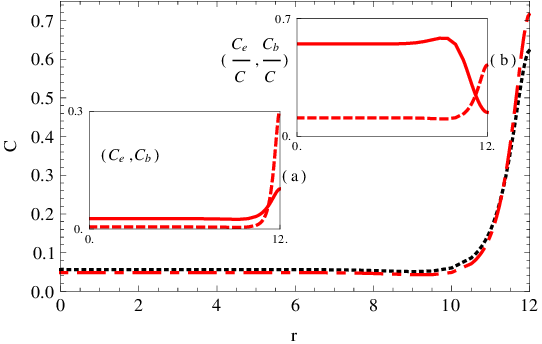}
  \caption{\label{candu} Concentration profile for \textit{N=1000} with $\mu$ corresponding to salt
    concentration 100 $mM$ and two different branching fugacities,
    $f_b=0$ (linear polymer) for the dotted line, and $f_b=3.0$
    (branched polymer) for the dotted-dashed line.  Inset (a) shows
    the concentration profile of endpoints (solid line) and branch points (dashed lines).
     Inset (b) shows the fraction of endpoints (solid line) and branch points (dashed lines).
    Other parameters used are $\upsilon=0.5$,
    $\tau=-1$, $\sigma=0.4$, $b=12$, and $T=300 K$.}
\end{figure}

The overall dimensionless monomer density profiles
$\mathcal{C}(r)=\Psi(r)^2$ as a function $r = |\textbf{r}|$ the distance from
the center of the cavity are shown in Fig.~\ref{candu} for a linear
polymer with $f_b=0$ and a branched polymer with $f_b=3.0$ of an equal
number of segments, $N = 1000$, enclosed in a spherical shell. The
radius of the capsid is taken to be $b = 12$ in units of the polymer
Kuhn length that ftor our purpose is of the order of 1 nm \cite{Vanderschoot2009}.
Both types of polymers can adsorb onto the surface and, interestingly, the
branched polymer is adsorbed more strongly onto the surface than the linear chain.

We also investigated the spatial inhomogeneity in our annealed
branched polymer model of RNA.  In Fig.~\ref{candu} (a), we
plot the dimensionless density of endpoints $C_e(r)=f_e\Psi(r)$ (solid line) and branches $C_b(r) =
\frac{f_b}{6}\Psi^3(r)$ (dashed lines), obtained from Eq.~\eqref{NeNb}. Figure~\ref{candu} (b) 
illustrates the fractions of end points $C_e/C$ (solid line)
and fraction of branches $C_b/C$ (dashed lines) as a
function of the distance from the center of the capsid.  We observe
that most branching takes place within a two Debye length layer, thus very near the capsid wall
where the concentration of segments is high and the
local gradient in density is the largest. This is straightforward to understand as branching
increases the local density allowing more segments to interact with
the wall. Figure \ref{candu} also shows that end points are dominantly distributed over the capsid interior. 
Thus branching can affect not only
the segment distribution but also the structure of the adsorbed layer making both quite
non-uniform.

 If we insert the parameter $f_e$ and fields $\Psi$ and $\Phi$
 numerically obtained from Eqs.~\eqref{euler},
 Eq.~\eqref{branch_constraint}, and Eq.~\eqref{constraint} into
 Eq.~\eqref{free_energy}, we can calculate the free energy of the
 chain-capsid complex, $F$.  To obtain the encapsulation free energy using Eq.~\ref{free_energy},
 we need to find the free energy difference between the chain-capsid complex and a free 
 chain in solution and a positively charged capsid. The free energy of the free self-interacting 
 chains (both linear and branched) is negligible under the stated 
 conditions and is ignored. The capsid self-energy resulting from electrostatic interaction is
 calculated solving the system in the limit as $N \to 0$. Obviously, the capsid self-energy
 does not depend on the genome topology, but is not negligible and is explicitly subtracted. 
 Here, we emphasize that since capsid proteins spontaneously self-assemble in the absence of genome in different kind of viruses,
 we only focused on the free energy of complexation of chain-capsid interaction.

A plot of the encapsulation free energy $F$ vs the monomer number $N$, as shown in
Fig. \ref{fenergy}, confirms that the free energy minimum moves
towards longer chains, {\it i.e.}, allowing more monomers to be
encapsulated into the viral shell. In the inset to Fig.~\ref{fenergy},
we plot the optimal polymer length $N_{\text{min}}$ (defined as the
position of the free energy minimum) versus the branch point fugacity
$f_b$ for $\mu$ corresponding to 100 $mM$.  As illustrated in
Fig.~\ref{fenergy}, this effect is more pronounced at high salt
concentration.

For low salt concentrations, electrostatics overwhelms all the other interactions and branching becomes less pronounced, but
it still has an effect so that a branched polymer is packaged more efficiently than a linear polymer. 
For instance, at 10 $mM$ salt the free energy has a minimum around N=638 for the linear polymer
and N=773 for the branched polymer, see Fig.~\ref{fenergy}. At 100 $mM$ salt,  the optimal number of monomers for
the linear polymer is N = 534, but for the branched polymer increases by more than two times to N=1211.
The arrow on the $N$ axis of Fig.~\ref{fenergy} corresponds to $N=4\pi b^2 \sigma$, representing a neutral system where the number of
positive charges on the capsid wall is equal to the number of negative
charges on the polymer chain. The aforementioned results correspond to the Neumann BC. As noted above, 
our conclusions do not depend on the type of BCs (Dirichlet vs. Neumann) that we employed. For the Dirichlet BC, at 100 $mM$, 
salt the optimal number of monomers is 202 for a 
linear polymer, and 930 for a branched polymer, consistent with the results for the Neumann BC.

This observation reveals that overcharging in viral particles could be a direct result of the secondary,
i.e., branched, structure of viral RNA. We emphasize here that we repeated the above calculations for different surface 
charge densities, 
relevant to different viral capsids ($0.3 \leq \sigma \leq 0.9$), and found that for all cases, the number of charges 
on linear polymers is 
less than the number of positive charges on the capsid. Quite interestingly, we also found that for any given linear 
charge density of the chain, 
the optimal length of encapsulated branched
polymers is always larger than that of linear polymers.

Figure \ref{fenergy} also reveals the second important point, {\sl viz.} that the free energies associated with
branched polymers have deeper minima than those for linear
polymers for a set of salt concentrations. This effect explains why
some RNAs are encapsulated more efficiently than other RNAs, or other
linear polyelectrolytes for that matter.

If coupling between RNA branching and electrostatics represents a
robust mechanism, the details of its description should not be qualitatively
important. To this end, it is interesting to compare our results for
encapsulated charged branched polymer based on a field theoretic {\sl
Ansatz} for the statistics of branched polymers \cite{Lubensky} with
a very simple model with short-ranged attractive interaction between
different chain segments mimicking the self-pairing of RNA bases \cite{Borukhov}. We
consider a linear polymer and now define the
$W[\Psi]$ term as
\begin{equation}
  W[\Psi] = \frac{1}{2}(v-s w) \Psi^4 + u \Psi^6,
\end{equation}
with $s$ the average fraction of base pairs and $w$ the binding
energy. We also include the next term in the virial expansion in order
to stabilize the free energy since the total coefficient in front of
the $\psi^4$ term can become negative.  Calculating $F$ vs. $N$ curves
for increasing values of $s$, the average fraction of self-paired bases,
we find the same qualitative behavior as for increasing branching
fugacity: the position of the minimum moves towards longer
polymers (larger $N$) and the depth of the minimum increases
for increasing $s$.
{{For example, at 10 mM the free energy minimum is located at $N=632$ for s=0 and at $N=740$ for $s=0.04$. 
At 100 mM salt the location of minimum moves from N=524 for s=0 to N=903 for s=0.04.}}  
Furthermore, as was the case for the branching
model, the free energy minimum becomes deeper as the number of base-pairs increases.

\section{DISCUSSION AND CONCLUSIONS}
\vspace{-0.25cm}

Both models described above show that the total charge of the genome
inside the capsid is larger than the one residing on the capsid
interior and that the virion is thus overcharged. Our analysis clearly
reveals that the genomic function of RNA, as encoded in its sequence
that engenders its branched secondary structure, plays an important
role in the self-assembly of ssRNA viruses. The branched secondary
structure of RNA, treated with either branching or self-pairing
models, promotes overcharging of the virion and stabilize its
equilibrium configuration. {We emphasize that while our results differ from previous studies \cite{Ting}, 
a very recent coarse-grained molecular dynamics simulation of the assembly of viral particles completely
confirms the importance of the structure of RNA in the assembly process \cite{Hagan}}. Note that within our field theory formalism, 
we do not observe overcharging for linear chains.  The condition of  the solution and protein charge distribution in the simulations 
of Ref. \cite{Hagan} are such that overcharging could be observed for linear chains; nevertheless, polymer branching 
enhances overcharging, consistent with our studies.

In order to explain the experiments noted in the introduction on the competition
between RNA of CCMV and BMV through the theory presented above, we calculated the number of branch points for both RNAs. 
In particular, we used RNASubopt, a program in the Vienna RNA package \cite{Vienna}, to generate an ensemble of secondary structures 
for genome sequences of RNA1 of BMV and CCMV.  We then
calculated the thermally averaged number of branch points from the
secondary structures of each RNA.  We found that RNA1 of BMV has higher average number of branch points (65) than CCMV
(60.5) confirming that in the absence of specific interactions RNA1 of
BMV would be preferentially packaged over RNA1 of CCMV, consistent with the
experimental results of Comas-Garcia et al. \cite{Comas,standard}. While one
has to be cautious about results for longer sequences {{at high salt concentrations}}, the
Vienna RNA Package \cite{Vienna} has been used to calculate thermally averaged properties
of viral genomes with lengths of 2500-7000 nt, and important results
have been obtained \cite{Yoffe2008}.

A comprehensive investigation of the physico-chemical
parameters that impact capsid formation could have great potential in
the development of anti-viral therapies and a systematic understanding of
the processes involved in viral infection.

\section*{ACKNOWLEDGMENTS}
\vspace{-0.25cm}

The authors would like to thank Mehran Kardar and Aaron Yoffe for useful
discussions. This work was supported by the National Science
Foundation through Grant No. DMR-13-10687 (R.Z.).

\bibliography{prl_revised_final_arxiv}

\begin{thebibliography}{42}
\expandafter\ifx\csname natexlab\endcsname\relax\def\natexlab#1{#1}\fi
\expandafter\ifx\csname bibnamefont\endcsname\relax
  \def\bibnamefont#1{#1}\fi
\expandafter\ifx\csname bibfnamefont\endcsname\relax
  \def\bibfnamefont#1{#1}\fi
\expandafter\ifx\csname citenamefont\endcsname\relax
  \def\citenamefont#1{#1}\fi
\expandafter\ifx\csname url\endcsname\relax
  \def\url#1{\texttt{#1}}\fi
\expandafter\ifx\csname urlprefix\endcsname\relax\def\urlprefix{URL }\fi
\providecommand{\bibinfo}[2]{#2}
\providecommand{\eprint}[2][]{\url{#2}}

\bibitem[{\citenamefont{Bancroft}(1970)}]{Bancroft}
\bibinfo{author}{\bibfnamefont{J.~B.} \bibnamefont{Bancroft}},
  \bibinfo{journal}{Adv. Virus Res.} \textbf{\bibinfo{volume}{16}},
  \bibinfo{pages}{99} (\bibinfo{year}{1970}).

\bibitem[{\citenamefont{Bruinsma}(2006)}]{bruinsma}
\bibinfo{author}{\bibfnamefont{R.~F.} \bibnamefont{Bruinsma}},
  \bibinfo{journal}{Euro. Phys. J. E} \textbf{\bibinfo{volume}{19}},
  \bibinfo{pages}{303} (\bibinfo{year}{2006}).

\bibitem[{\citenamefont{Sikkema et~al.}(2007)\citenamefont{Sikkema,
  Comellas-Aragones, Fokkink, Verduin, Cornelissen, and
  Nolte}}]{Cornelissen2007}
\bibinfo{author}{\bibfnamefont{F.~D.} \bibnamefont{Sikkema}},
  \bibinfo{author}{\bibfnamefont{M.}~\bibnamefont{Comellas-Aragones}},
  \bibinfo{author}{\bibfnamefont{R.~G.} \bibnamefont{Fokkink}},
  \bibinfo{author}{\bibfnamefont{B.~J.~M.} \bibnamefont{Verduin}},
  \bibinfo{author}{\bibfnamefont{J.}~\bibnamefont{Cornelissen}},
  \bibnamefont{and} \bibinfo{author}{\bibfnamefont{R.~J.~M.}
  \bibnamefont{Nolte}}, \bibinfo{journal}{Org. Biomol. Chem.}
  \textbf{\bibinfo{volume}{5}}, \bibinfo{pages}{54} (\bibinfo{year}{2007}).

\bibitem[{\citenamefont{Ren et~al.}(2006)\citenamefont{Ren, Wong, and
  Lim}}]{Ren2006}
\bibinfo{author}{\bibfnamefont{Y.~P.} \bibnamefont{Ren}},
  \bibinfo{author}{\bibfnamefont{S.~M.} \bibnamefont{Wong}}, \bibnamefont{and}
  \bibinfo{author}{\bibfnamefont{L.~Y.} \bibnamefont{Lim}},
  \bibinfo{journal}{J. Gen. Virol.} \textbf{\bibinfo{volume}{87}},
  \bibinfo{pages}{2749} (\bibinfo{year}{2006}).

\bibitem[{\citenamefont{Ni et~al.}(2012)\citenamefont{Ni, Wang, Ma, Das, Sokol,
  Chiu, Dragnea, Hagan, and Kao}}]{Bogdan}
\bibinfo{author}{\bibfnamefont{P.}~\bibnamefont{Ni}},
  \bibinfo{author}{\bibfnamefont{Z.}~\bibnamefont{Wang}},
  \bibinfo{author}{\bibfnamefont{X.}~\bibnamefont{Ma}},
  \bibinfo{author}{\bibfnamefont{N.~C.} \bibnamefont{Das}},
  \bibinfo{author}{\bibfnamefont{P.}~\bibnamefont{Sokol}},
  \bibinfo{author}{\bibfnamefont{W.}~\bibnamefont{Chiu}},
  \bibinfo{author}{\bibfnamefont{B.}~\bibnamefont{Dragnea}},
  \bibinfo{author}{\bibfnamefont{M.}~\bibnamefont{Hagan}}, \bibnamefont{and}
  \bibinfo{author}{\bibfnamefont{C.~C.} \bibnamefont{Kao}},
  \bibinfo{journal}{J. Mol. Biol.} \textbf{\bibinfo{volume}{419}},
  \bibinfo{pages}{284} (\bibinfo{year}{2012}).

\bibitem[{\citenamefont{van~der Schoot and Zandi}(2007)}]{Vanderschoot2007}
\bibinfo{author}{\bibfnamefont{P.}~\bibnamefont{van~der Schoot}}
  \bibnamefont{and} \bibinfo{author}{\bibfnamefont{R.}~\bibnamefont{Zandi}},
  \bibinfo{journal}{Phys. Biol.} \textbf{\bibinfo{volume}{4}},
  \bibinfo{pages}{296} (\bibinfo{year}{2007}).

\bibitem[{\citenamefont{Losdorfer~Bozic
  et~al.}(2013)\citenamefont{Losdorfer~Bozic, Siber, and Podgornik}}]{Anze2}
\bibinfo{author}{\bibfnamefont{A.}~\bibnamefont{Losdorfer~Bozic}},
  \bibinfo{author}{\bibfnamefont{A.}~\bibnamefont{Siber}}, \bibnamefont{and}
  \bibinfo{author}{\bibfnamefont{R.}~\bibnamefont{Podgornik}},
  \bibinfo{journal}{J. Biol. Phys.} \textbf{\bibinfo{volume}{39}},
  \bibinfo{pages}{215} (\bibinfo{year}{2013}).

\bibitem[{\citenamefont{Daniel et~al.}(2010)\citenamefont{Daniel, Tsvetkova,
  Quinkert, Murali, De, Rotello, Kao, and Dragnea}}]{DanielDragnea2010}
\bibinfo{author}{\bibfnamefont{M.~C.} \bibnamefont{Daniel}},
  \bibinfo{author}{\bibfnamefont{I.~B.} \bibnamefont{Tsvetkova}},
  \bibinfo{author}{\bibfnamefont{Z.~T.} \bibnamefont{Quinkert}},
  \bibinfo{author}{\bibfnamefont{A.}~\bibnamefont{Murali}},
  \bibinfo{author}{\bibfnamefont{M.}~\bibnamefont{De}},
  \bibinfo{author}{\bibfnamefont{V.~M.} \bibnamefont{Rotello}},
  \bibinfo{author}{\bibfnamefont{C.~C.} \bibnamefont{Kao}}, \bibnamefont{and}
  \bibinfo{author}{\bibfnamefont{B.}~\bibnamefont{Dragnea}},
  \bibinfo{journal}{ACS Nano} \textbf{\bibinfo{volume}{4}},
  \bibinfo{pages}{3853} (\bibinfo{year}{2010}).

\bibitem[{\citenamefont{Zlotnick et~al.}(2000)\citenamefont{Zlotnick, Aldrich,
  Johnson, Ceres, and Young}}]{Zlotnick}
\bibinfo{author}{\bibfnamefont{A.}~\bibnamefont{Zlotnick}},
  \bibinfo{author}{\bibfnamefont{R.}~\bibnamefont{Aldrich}},
  \bibinfo{author}{\bibfnamefont{J.~M.} \bibnamefont{Johnson}},
  \bibinfo{author}{\bibfnamefont{P.}~\bibnamefont{Ceres}}, \bibnamefont{and}
  \bibinfo{author}{\bibfnamefont{M.~J.} \bibnamefont{Young}},
  \bibinfo{journal}{Virology} \textbf{\bibinfo{volume}{277}},
  \bibinfo{pages}{450} (\bibinfo{year}{2000}).

\bibitem[{\citenamefont{Lin et~al.}(2012)\citenamefont{Lin, van~der Schoot, and
  Zandi}}]{Hsiang-Ku}
\bibinfo{author}{\bibfnamefont{H.-K.} \bibnamefont{Lin}},
  \bibinfo{author}{\bibfnamefont{P.}~\bibnamefont{van~der Schoot}},
  \bibnamefont{and} \bibinfo{author}{\bibfnamefont{R.}~\bibnamefont{Zandi}},
  \bibinfo{journal}{Phys. Biol.} \textbf{\bibinfo{volume}{9}},
  \bibinfo{pages}{066004} (\bibinfo{year}{2012}).

\bibitem[{\citenamefont{Comas-Garcia et~al.}(2012)\citenamefont{Comas-Garcia,
  Cadena-Nava, Rao, Knobler, and Gelbart}}]{Comas}
\bibinfo{author}{\bibfnamefont{M.}~\bibnamefont{Comas-Garcia}},
  \bibinfo{author}{\bibfnamefont{R.~D.} \bibnamefont{Cadena-Nava}},
  \bibinfo{author}{\bibfnamefont{A.~L.~N.} \bibnamefont{Rao}},
  \bibinfo{author}{\bibfnamefont{C.~M.} \bibnamefont{Knobler}},
  \bibnamefont{and} \bibinfo{author}{\bibfnamefont{W.~M.}
  \bibnamefont{Gelbart}}, \bibinfo{journal}{J. Virol.}
  \textbf{\bibinfo{volume}{86}}, \bibinfo{pages}{12271} (\bibinfo{year}{2012}).

\bibitem[{\citenamefont{Yoffe et~al.}(2008)\citenamefont{Yoffe, Prinsen, Gopal,
  Knobler, Gelbart, and Ben-Shaul}}]{Yoffe2008}
\bibinfo{author}{\bibfnamefont{A.~M.} \bibnamefont{Yoffe}},
  \bibinfo{author}{\bibfnamefont{P.}~\bibnamefont{Prinsen}},
  \bibinfo{author}{\bibfnamefont{A.}~\bibnamefont{Gopal}},
  \bibinfo{author}{\bibfnamefont{C.~M.} \bibnamefont{Knobler}},
  \bibinfo{author}{\bibfnamefont{W.~M.} \bibnamefont{Gelbart}},
  \bibnamefont{and}
  \bibinfo{author}{\bibfnamefont{A.}~\bibnamefont{Ben-Shaul}},
  \bibinfo{journal}{PNAS} \textbf{\bibinfo{volume}{105}},
  \bibinfo{pages}{16153} (\bibinfo{year}{2008}).

\bibitem[{\citenamefont{Li~Tai et~al.}(2011)\citenamefont{Li~Tai, Gelbart, and
  Ben-Shaul}}]{Li-tai}
\bibinfo{author}{\bibfnamefont{F.}~\bibnamefont{Li~Tai}},
  \bibinfo{author}{\bibfnamefont{W.~M.} \bibnamefont{Gelbart}},
  \bibnamefont{and}
  \bibinfo{author}{\bibfnamefont{A.}~\bibnamefont{Ben-Shaul}},
  \bibinfo{journal}{J. Chem. Phys.} \textbf{\bibinfo{volume}{135}},
  \bibinfo{pages}{155105} (\bibinfo{year}{2011}).

\bibitem[{\citenamefont{Tao et~al.}(2008)\citenamefont{Tao, Rui, and
  Shklovskii}}]{Shklovskii}
\bibinfo{author}{\bibfnamefont{H.}~\bibnamefont{Tao}},
  \bibinfo{author}{\bibfnamefont{Z.}~\bibnamefont{Rui}}, \bibnamefont{and}
  \bibinfo{author}{\bibfnamefont{B.~I.} \bibnamefont{Shklovskii}},
  \bibinfo{journal}{Physica A} \textbf{\bibinfo{volume}{387}},
  \bibinfo{pages}{3059} (\bibinfo{year}{2008}).

\bibitem[{\citenamefont{van~der Schoot and Bruinsma}(2005)}]{Vanderschoot}
\bibinfo{author}{\bibfnamefont{P.}~\bibnamefont{van~der Schoot}}
  \bibnamefont{and} \bibinfo{author}{\bibfnamefont{R.}~\bibnamefont{Bruinsma}},
  \bibinfo{journal}{Phys. Rev. E} \textbf{\bibinfo{volume}{71}},
  \bibinfo{pages}{061928} (\bibinfo{year}{2005}).

\bibitem[{\citenamefont{Belyi and Muthukumar}(2006)}]{Belyi}
\bibinfo{author}{\bibfnamefont{V.~A.} \bibnamefont{Belyi}} \bibnamefont{and}
  \bibinfo{author}{\bibfnamefont{M.}~\bibnamefont{Muthukumar}},
  \bibinfo{journal}{PNAS} \textbf{\bibinfo{volume}{103}},
  \bibinfo{pages}{17174} (\bibinfo{year}{2006}).

\bibitem[{\citenamefont{Ting et~al.}(2011)\citenamefont{Ting, Wu, and
  Wang}}]{Ting}
\bibinfo{author}{\bibfnamefont{C.~L.} \bibnamefont{Ting}},
  \bibinfo{author}{\bibfnamefont{J.~Z.} \bibnamefont{Wu}}, \bibnamefont{and}
  \bibinfo{author}{\bibfnamefont{Z.~G.} \bibnamefont{Wang}},
  \bibinfo{journal}{PNAS} \textbf{\bibinfo{volume}{108}},
  \bibinfo{pages}{16986} (\bibinfo{year}{2011}).

\bibitem[{\citenamefont{Hu et~al.}(2008)\citenamefont{Hu, Zandi, Anavitarte,
  Knobler, and Gelbart}}]{Chuck2008}
\bibinfo{author}{\bibfnamefont{Y.~F.} \bibnamefont{Hu}},
  \bibinfo{author}{\bibfnamefont{R.}~\bibnamefont{Zandi}},
  \bibinfo{author}{\bibfnamefont{A.}~\bibnamefont{Anavitarte}},
  \bibinfo{author}{\bibfnamefont{C.~M.} \bibnamefont{Knobler}},
  \bibnamefont{and} \bibinfo{author}{\bibfnamefont{W.~M.}
  \bibnamefont{Gelbart}}, \bibinfo{journal}{Biophys. J.}
  \textbf{\bibinfo{volume}{94}}, \bibinfo{pages}{1428} (\bibinfo{year}{2008}).

\bibitem[{\citenamefont{Cadena-Nava et~al.}(2011)\citenamefont{Cadena-Nava, Hu,
  Garmann, Ng, Zelikin, Knobler, and Gelbart}}]{Cadena2011}
\bibinfo{author}{\bibfnamefont{R.~D.} \bibnamefont{Cadena-Nava}},
  \bibinfo{author}{\bibfnamefont{Y.~F.} \bibnamefont{Hu}},
  \bibinfo{author}{\bibfnamefont{R.~F.} \bibnamefont{Garmann}},
  \bibinfo{author}{\bibfnamefont{B.}~\bibnamefont{Ng}},
  \bibinfo{author}{\bibfnamefont{A.~N.} \bibnamefont{Zelikin}},
  \bibinfo{author}{\bibfnamefont{C.~M.} \bibnamefont{Knobler}},
  \bibnamefont{and} \bibinfo{author}{\bibfnamefont{W.~M.}
  \bibnamefont{Gelbart}}, \bibinfo{journal}{J. Phys. Chem. B}
  \textbf{\bibinfo{volume}{115}}, \bibinfo{pages}{2386} (\bibinfo{year}{2011}).

\bibitem[{\citenamefont{Zandi and van~der Schoot}(2009)}]{Vanderschoot2009}
\bibinfo{author}{\bibfnamefont{R.}~\bibnamefont{Zandi}} \bibnamefont{and}
  \bibinfo{author}{\bibfnamefont{P.}~\bibnamefont{van~der Schoot}},
  \bibinfo{journal}{Biophys. J.} \textbf{\bibinfo{volume}{96}},
  \bibinfo{pages}{9} (\bibinfo{year}{2009}).

\bibitem[{\citenamefont{Bozic et~al.}(2012)\citenamefont{Bozic, Siber, and
  Podgornik}}]{Anze}
\bibinfo{author}{\bibfnamefont{A.~L.} \bibnamefont{Bozic}},
  \bibinfo{author}{\bibfnamefont{A.}~\bibnamefont{Siber}}, \bibnamefont{and}
  \bibinfo{author}{\bibfnamefont{R.}~\bibnamefont{Podgornik}},
  \bibinfo{journal}{J. Biol. Phys.} \textbf{\bibinfo{volume}{38}},
  \bibinfo{pages}{657} (\bibinfo{year}{2012}).

\bibitem[{\citenamefont{Siber and Podgornik}(2008)}]{Siber}
\bibinfo{author}{\bibfnamefont{A.}~\bibnamefont{Siber}} \bibnamefont{and}
  \bibinfo{author}{\bibfnamefont{R.}~\bibnamefont{Podgornik}},
  \bibinfo{journal}{Phys. Rev. E} \textbf{\bibinfo{volume}{78}},
  \bibinfo{pages}{051915} (\bibinfo{year}{2008}).

\bibitem[{\citenamefont{Elrad and Hagan}(2010)}]{Hagan2010}
\bibinfo{author}{\bibfnamefont{O.~M.} \bibnamefont{Elrad}} \bibnamefont{and}
  \bibinfo{author}{\bibfnamefont{M.~F.} \bibnamefont{Hagan}},
  \bibinfo{journal}{Phys. Biol.} \textbf{\bibinfo{volume}{7}},
  \bibinfo{pages}{045003} (\bibinfo{year}{2010}).

\bibitem[{\citenamefont{Hagan}(2014)}]{Hagan2013}
\bibinfo{author}{\bibfnamefont{M.~F.} \bibnamefont{Hagan}},
  \bibinfo{journal}{Adv. Chem. Phys.} \textbf{\bibinfo{volume}{155}},
  \bibinfo{pages}{1} (\bibinfo{year}{2014}).

\bibitem[{\citenamefont{Borodavka et~al.}(2012)\citenamefont{Borodavka, Tuma,
  and Stockley}}]{Stockley2012}
\bibinfo{author}{\bibfnamefont{A.}~\bibnamefont{Borodavka}},
  \bibinfo{author}{\bibfnamefont{R.}~\bibnamefont{Tuma}}, \bibnamefont{and}
  \bibinfo{author}{\bibfnamefont{P.~G.} \bibnamefont{Stockley}},
  \bibinfo{journal}{PNAS} \textbf{\bibinfo{volume}{109}},
  \bibinfo{pages}{15769} (\bibinfo{year}{2012}).

\bibitem[{\citenamefont{Stockley et~al.}(2013)\citenamefont{Stockley, Twarock,
  Bakker, Barker, Borodavka, Dykeman, Ford, Pearson, Phillips, Ranson
  et~al.}}]{Stockley2013}
\bibinfo{author}{\bibfnamefont{P.~G.} \bibnamefont{Stockley}},
  \bibinfo{author}{\bibfnamefont{R.}~\bibnamefont{Twarock}},
  \bibinfo{author}{\bibfnamefont{S.~E.} \bibnamefont{Bakker}},
  \bibinfo{author}{\bibfnamefont{A.~M.} \bibnamefont{Barker}},
  \bibinfo{author}{\bibfnamefont{A.}~\bibnamefont{Borodavka}},
  \bibinfo{author}{\bibfnamefont{E.}~\bibnamefont{Dykeman}},
  \bibinfo{author}{\bibfnamefont{R.~J.} \bibnamefont{Ford}},
  \bibinfo{author}{\bibfnamefont{A.~R.} \bibnamefont{Pearson}},
  \bibinfo{author}{\bibfnamefont{S.~E.~V.} \bibnamefont{Phillips}},
  \bibinfo{author}{\bibfnamefont{N.~A.} \bibnamefont{Ranson}},
  \bibnamefont{et~al.}, \bibinfo{journal}{J. Biol. Phys.}
  \textbf{\bibinfo{volume}{39}}, \bibinfo{pages}{277} (\bibinfo{year}{2013}).

\bibitem[{\citenamefont{McPherson}(2005)}]{McPherson}
\bibinfo{author}{\bibfnamefont{A.}~\bibnamefont{McPherson}},
  \bibinfo{journal}{BioEssays} \textbf{\bibinfo{volume}{27}},
  \bibinfo{pages}{447} (\bibinfo{year}{2005}).

\bibitem[{\citenamefont{Borukhov et~al.}(1998)\citenamefont{Borukhov, Andelman,
  and Orland}}]{Borukhov}
\bibinfo{author}{\bibfnamefont{I.}~\bibnamefont{Borukhov}},
  \bibinfo{author}{\bibfnamefont{D.}~\bibnamefont{Andelman}}, \bibnamefont{and}
  \bibinfo{author}{\bibfnamefont{H.}~\bibnamefont{Orland}},
  \bibinfo{journal}{Euro. Phys. J. B} \textbf{\bibinfo{volume}{5}},
  \bibinfo{pages}{869} (\bibinfo{year}{1998}).

\bibitem[{\citenamefont{Borukhov et~al.}(1995)\citenamefont{Borukhov, Andelman,
  and Orland}}]{Borukhov1}
\bibinfo{author}{\bibfnamefont{I.}~\bibnamefont{Borukhov}},
  \bibinfo{author}{\bibfnamefont{D.}~\bibnamefont{Andelman}}, \bibnamefont{and}
  \bibinfo{author}{\bibfnamefont{H.}~\bibnamefont{Orland}},
  \bibinfo{journal}{Europhys. Lett.} \textbf{\bibinfo{volume}{32}},
  \bibinfo{pages}{499} (\bibinfo{year}{1995}).

\bibitem[{\citenamefont{Borukhov and Andelman}(1998)}]{Borukhov2}
\bibinfo{author}{\bibfnamefont{I.}~\bibnamefont{Borukhov}} \bibnamefont{and}
  \bibinfo{author}{\bibfnamefont{D.}~\bibnamefont{Andelman}},
  \bibinfo{journal}{Macromolecules} \textbf{\bibinfo{volume}{31}},
  \bibinfo{pages}{1665} (\bibinfo{year}{1998}).

\bibitem[{\citenamefont{Borukhov et~al.}(1999)\citenamefont{Borukhov, Andelman,
  and Orland}}]{Borukhov3}
\bibinfo{author}{\bibfnamefont{I.}~\bibnamefont{Borukhov}},
  \bibinfo{author}{\bibfnamefont{D.}~\bibnamefont{Andelman}}, \bibnamefont{and}
  \bibinfo{author}{\bibfnamefont{H.}~\bibnamefont{Orland}},
  \bibinfo{journal}{J. Phys. Chem. B} \textbf{\bibinfo{volume}{103}},
  \bibinfo{pages}{5042} (\bibinfo{year}{1999}).

\bibitem[{\citenamefont{Shafir et~al.}(2003)\citenamefont{Shafir, Andelman, and
  Netz}}]{Shafir}
\bibinfo{author}{\bibfnamefont{A.}~\bibnamefont{Shafir}},
  \bibinfo{author}{\bibfnamefont{D.}~\bibnamefont{Andelman}}, \bibnamefont{and}
  \bibinfo{author}{\bibfnamefont{R.~R.} \bibnamefont{Netz}},
  \bibinfo{journal}{J. Chem. Phys.} \textbf{\bibinfo{volume}{119}},
  \bibinfo{pages}{2355} (\bibinfo{year}{2003}).

\bibitem[{\citenamefont{Lubensky and Isaacson}(1979)}]{Lubensky}
\bibinfo{author}{\bibfnamefont{T.~C.} \bibnamefont{Lubensky}} \bibnamefont{and}
  \bibinfo{author}{\bibfnamefont{J.}~\bibnamefont{Isaacson}},
  \bibinfo{journal}{Phys. Rev. A} \textbf{\bibinfo{volume}{20}},
  \bibinfo{pages}{2130} (\bibinfo{year}{1979}).

\bibitem[{\citenamefont{Nguyen and Bruinsma}(2006)}]{Nguyen-Bruinsma}
\bibinfo{author}{\bibfnamefont{T.~T.} \bibnamefont{Nguyen}} \bibnamefont{and}
  \bibinfo{author}{\bibfnamefont{R.~F.} \bibnamefont{Bruinsma}},
  \bibinfo{journal}{Phys. Rev. Lett.} \textbf{\bibinfo{volume}{97}},
  \bibinfo{pages}{108102} (\bibinfo{year}{2006}).

\bibitem[{\citenamefont{Lee and Nguyen}(2008)}]{Lee-Nguyen}
\bibinfo{author}{\bibfnamefont{S.~I.} \bibnamefont{Lee}} \bibnamefont{and}
  \bibinfo{author}{\bibfnamefont{T.~T.} \bibnamefont{Nguyen}},
  \bibinfo{journal}{Phys. Rev. Lett.} \textbf{\bibinfo{volume}{100}},
  \bibinfo{pages}{198102} (\bibinfo{year}{2008}).

\bibitem[{\citenamefont{Elleuch et~al.}(1995)\citenamefont{Elleuch, Lequeux,
  and Pfeuty}}]{Elleuch}
\bibinfo{author}{\bibfnamefont{K.}~\bibnamefont{Elleuch}},
  \bibinfo{author}{\bibfnamefont{F.}~\bibnamefont{Lequeux}}, \bibnamefont{and}
  \bibinfo{author}{\bibfnamefont{P.}~\bibnamefont{Pfeuty}},
  \bibinfo{journal}{J. Phys. I France} \textbf{\bibinfo{volume}{5}},
  \bibinfo{pages}{465} (\bibinfo{year}{1995}).

\bibitem[{\citenamefont{de~Gennes}(1982)}]{deGennes}
\bibinfo{author}{\bibfnamefont{P.-G.} \bibnamefont{de~Gennes}},
  \bibinfo{journal}{Macromolecules} \textbf{\bibinfo{volume}{15}},
  \bibinfo{pages}{492} (\bibinfo{year}{1982}).

\bibitem[{\citenamefont{Ji and Hone}(1988)}]{Hone}
\bibinfo{author}{\bibfnamefont{H.}~\bibnamefont{Ji}} \bibnamefont{and}
  \bibinfo{author}{\bibfnamefont{D.}~\bibnamefont{Hone}},
  \bibinfo{journal}{Macromolecules} \textbf{\bibinfo{volume}{21}},
  \bibinfo{pages}{2600} (\bibinfo{year}{1988}).

\bibitem[{\citenamefont{Siber et~al.}(2010)\citenamefont{Siber, Zandi, and
  Podgornik}}]{SiberZandi2010}
\bibinfo{author}{\bibfnamefont{A.}~\bibnamefont{Siber}},
  \bibinfo{author}{\bibfnamefont{R.}~\bibnamefont{Zandi}}, \bibnamefont{and}
  \bibinfo{author}{\bibfnamefont{R.}~\bibnamefont{Podgornik}},
  \bibinfo{journal}{Phys. Rev. E} \textbf{\bibinfo{volume}{81}},
  \bibinfo{pages}{051919} (\bibinfo{year}{2010}).

\bibitem[{\citenamefont{Perlmutter et~al.}(2013)\citenamefont{Perlmutter, Qiao,
  and Hagan}}]{Hagan}
\bibinfo{author}{\bibfnamefont{J.~D.} \bibnamefont{Perlmutter}},
  \bibinfo{author}{\bibfnamefont{C.}~\bibnamefont{Qiao}}, \bibnamefont{and}
  \bibinfo{author}{\bibfnamefont{M.~F.} \bibnamefont{Hagan}},
  \bibinfo{journal}{eLife} \textbf{\bibinfo{volume}{2}} (\bibinfo{year}{2013}).

\bibitem[{\citenamefont{Hofacker et~al.}(1994)\citenamefont{Hofacker, Fontana,
  Stadler, Bonhoeffer, Tacker, and Schuster}}]{Vienna}
\bibinfo{author}{\bibfnamefont{I.~L.} \bibnamefont{Hofacker}},
  \bibinfo{author}{\bibfnamefont{W.}~\bibnamefont{Fontana}},
  \bibinfo{author}{\bibfnamefont{P.~F.} \bibnamefont{Stadler}},
  \bibinfo{author}{\bibfnamefont{L.~S.} \bibnamefont{Bonhoeffer}},
  \bibinfo{author}{\bibfnamefont{M.}~\bibnamefont{Tacker}}, \bibnamefont{and}
  \bibinfo{author}{\bibfnamefont{P.}~\bibnamefont{Schuster}},
  \bibinfo{journal}{Monatsh. Chem.} \textbf{\bibinfo{volume}{125}},
  \bibinfo{pages}{167} (\bibinfo{year}{1994}).

\bibitem[{sta()}]{standard}
\bibinfo{note}{The standard deviation for the number of branch points in the
  ensemble of RNA secondary structures is around 2.5 for the RNA1 sequence of
  both BMV and CCMV.}

\end{thebibliography}
\end{document}